\documentclass[12]{iopart}
\usepackage{graphicx}
\usepackage{hyperref}
\usepackage{amsmathred}
\usepackage{amsfonts}
\usepackage{amssymb}
\usepackage{dcolumn}
\usepackage{bm}
\usepackage{epsfig}
\usepackage{color}
\usepackage{times,txfonts}
\newcommand{\bra}[1]{\langle#1|}

\newcommand{\ket}[1]{|#1\rangle}

\begin{document}

\title[Hierarchy and dynamics of trace distance correlations]{{\sf \bfseries Hierarchy and dynamics of trace distance correlations}}

\author{{\sf \bfseries Benjamin Aaronson$^1$,  Rosario Lo Franco$^2$, Giuseppe Compagno$^2$, and Gerardo Adesso$^1$}}
\address{$^1$School of Mathematical Sciences, The University of Nottingham, University Park, Nottingham NG7 2RD, United Kingdom}
\address{$^2$Dipartimento di Fisica e Chimica, Universit\`a di Palermo, via Archirafi 36, 90123 Palermo,
Italy}
\pacs{03.65.Ud, 03.65.Ta, 03.67.Ac, 03.67.Mn}

\begin{abstract}
We define and analyze measures of correlations for bipartite states based on trace distance. For Bell diagonal states of two qubits, in addition to the known expression for quantum correlations using this metric, we provide analytic expressions for the classical and total correlations. The ensuing hierarchy of correlations based on trace distance  is compared to the ones based on relative entropy and Hilbert-Schmidt norm. Although some common features can be found, the trace distance measure is shown to differentiate from the others in that the closest uncorrelated state to a given bipartite quantum state is not given by the product of the marginals, and further, the total correlations are strictly smaller than the sum of the quantum and classical correlations. We compare the various correlation measures in two dynamical non-Markovian models, locally applied phase-flip channels and random external fields. It is shown that the freezing behavior, observed across all known valid measures of quantum correlations for Bell diagonal states under local phase-flip channels, occurs for a larger set of starting states for the trace distance than for the other metrics.
\end{abstract}

\date{}

\maketitle

\tableofcontents
\title[Hierarchy and dynamics of trace distance correlations]
\smallskip
\vspace*{-1.5cm}
\section{Introduction}

Quantum entanglement is a central subject in the study of quantum information theory as it is a strikingly non-classical phenomenon and a primary instance of a  truly quantum resource in communication and computation tasks \cite{nielsenchuang,horodecki2009RMP}. However, in mixed states of composite systems, more general quantifiers of quantum correlations exist, most famously the quantum discord \cite{Zurek2001PRL,Henderson2001JPA}. Discord is present in most mixed states, even among those with no entanglement \cite{Ferraro2010PRA}, and it is of ongoing interest to investigate whether states with discord can be employed as resources for information processing scenarios \cite{modirev,streltsov,acti,lqu}, including those with vanishing entanglement \cite{Datta2008PRL}.

Some measures of quantum correlations, including the original discord \cite{Zurek2001PRL} and the one-way quantum deficit (alias relative entropy of discord)  \cite{Modi2010PRL,deficit}, are based on entropic quantities. Another method, the `geometric' approach for quantifying quantum correlations, consists in choosing a metric over the space of quantum states, and using this to find the distance to the nearest zero-discord (classical) state. Several measures have been defined in this way, including the Hilbert-Schmidt measure of discord \cite{dakic2010,luofu} and its modifications \cite{bellomo2012PRA,tufodiscord,luohell}. The trace distance measure of quantum correlations \cite{noq,sarandy} falls into the latter category.

The trace distance between two quantum states $\rho$ and $\sigma$ is defined as
\begin{equation}\label{TD}
\delta_\mathrm{TD}(\rho,\sigma)\equiv\frac12 \|\rho-\sigma\|_1\,,
 \end{equation}
where $\|\hat{O}\|_1\equiv\mathrm{Tr}|\hat{O}|=\mathrm{Tr}\sqrt{\hat{O}^\dag \hat{O}}$ is the Schatten-$1$ norm, or trace norm, with $\hat{O}$ being an arbitrary operator. The trace distance metric arises naturally in quantum mechanics and admits an intuitive operational interpretation related to the probability of successfully distinguishing between two quantum states in a hypothesis testing scenario \cite{helstrom}. An important feature of the trace distance in dynamical contexts is its contractivity under trace preserving and completely positive maps \cite{ruskai}.
A closed expression for the trace distance discord has been obtained for Bell diagonal states of two qubits \cite{noq,sarandy} and more generally for $X$-shaped states of two qubits \cite{giovannetti}. The trace distance discord has been theoretically studied in dynamical conditions in Refs.~\cite{sarandydec,aaronsonPRA}, and experimentally investigated in a nuclear magnetic resonance two-qubit system under phase and amplitude damping channels \cite{isabela}. These findings naturally encourage one to exploit the trace distance  to  introduce total and classical correlations as well, in order to construct a unified view of the correlations present in a composite quantum system and investigate their hierarchies and dynamical properties.

\begin{figure}[tb]
\begin{center}
{\includegraphics[width=8cm]{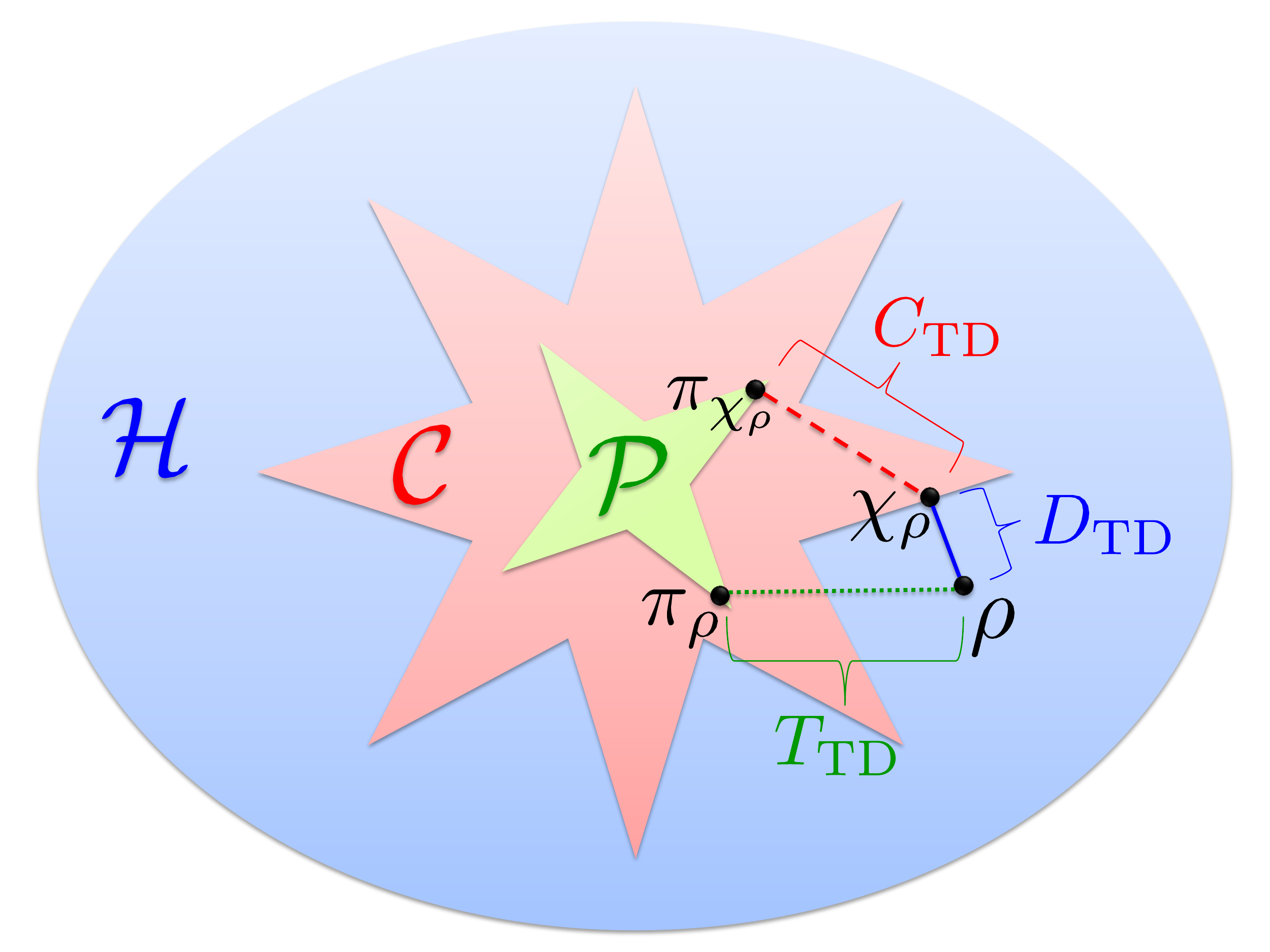}
\caption{\label{fipalla}(Color online) Schematic picture of the correlation hierarchy based on trace distance.
Given a state $\rho$ living in a bipartite Hilbert space ${\cal H}$, the
trace distance between $\rho$ and its closest classical state
$\chi_{\rho} \in {\cal P}$ defines the quantum correlations (discord)
$D_\mathrm{TD}$ of $\rho$. The trace distance between $\chi_{\rho}$ and
its closest product state $\pi_{\chi_{\rho}} \in {\cal P}$ defines the
classical correlations $C_\mathrm{TD}$  of $\rho$. The trace distance
between $\rho$ and its closest product state $\pi_\rho \in {\cal P}$
defines the total correlations $T_\mathrm{TD}$ of $\rho$. See Equations
(\ref{DTD}, \ref{TDTotal}, \ref{TDClassical}) in the main text for
rigorous definitions.}}
\end{center}
\end{figure}

In this paper, we construct a unified hierarchy of quantum, classical and total correlations in bipartite quantum states based on the trace distance (see Fig.~\ref{fipalla}). Unlike similar hierarchies based on relative entropy \cite{Modi2010PRL} or Hilbert-Schmidt norm \cite{bellomo2012PRA2}, the trace distance measures of correlations present surprising features. For Bell diagonal states of two qubits, we complement the study of \cite{noq,sarandy} by deriving closed expressions for the classical and total correlations  defined via trace distance. Counterintuitively, classical and quantum correlations do not add up to the total ones, not even for simple Bell diagonal states. In particular, the closest product state to a generic bipartite state, according to trace distance, is not in general the product of its marginals, which is instead the case e.g.~for relative entropy. We further investigate the dynamical evolution of quantum, classical and total correlations in typical non-Markovian environments by highlighting peculiar aspects and differences with the dynamics of relative entropy-based correlations.

This paper is organized as follows. In Section~\ref{sec:QuantumCorr}, we recall the expression, in the case of Bell diagonal states, for the trace distance measure of discord 
and we provide the explicit form of the associated closest classical states. In Section~\ref{sec:TotClassCorr}, we obtain expressions for the classical and total correlations of Bell diagonal states and discuss their features. In Section~\ref{sec:DynamQCorr}, we examine the behavior of quantum, classical, and total trace distance correlations in simple dynamical models. We conclude in Section~\ref{sec:Conc}.

\section{\label{sec:QuantumCorr}Quantum trace distance correlations and closest classical state}
We consider the class of two-qubit Bell diagonal states (or states with maximally mixed marginals \cite{Luo2008PRA}) expressed in the Bloch representation as
\begin{equation}\label{BelldiagonalstateBloch}
\rho_\mathrm{B}=\frac14\left(\mathbb{I}_A\otimes\mathbb{I}_B+\sum_{i=1}^3R_{ii}{\sigma_i}_A\otimes{\sigma_i}_B\right)\,,
\end{equation}
where the coefficients $R_{ii}$ are the nonzero correlation matrix elements and $\sigma_i$ are the Pauli matrices. In the basis of Bell states, a Bell diagonal state is instead written as
\begin{equation}\label{BelldiagonalstateBell}
\rho_\mathrm{B}=\sum_{j,r}\lambda_j^r\ket{j_r}\bra{j_r},\quad
\end{equation}
where $j=1,2$, $r=\pm$, $\sum_{j,r}\lambda_j^r=1$ and where we have indicated with $\ket{1_\pm}\equiv(\ket{01}\pm\ket{10})/\sqrt{2}$ the one-excitation Bell states and with $\ket{2_\pm}\equiv(\ket{00}\pm\ket{11})/\sqrt{2}$ the two-excitation Bell states. The relations among the eigenvalues $\lambda_j^r$ and the correlation matrix elements $R_{ii}$ are
\begin{equation}\label{lambdaR}
\lambda_1^\pm=\frac{1\pm R_{11}\pm R_{22} +R_{33}}{4},\quad
\lambda_2^\pm=\frac{1\pm R_{11}\mp R_{22} +R_{33}}{4},
\end{equation}
from which one obtains the inverse relations
\begin{eqnarray}\label{lambdaRinverse}
R_{11}&=&-1+2(\lambda_1^+ + \lambda_2^+),\nonumber\\
R_{22}&=&-1+2(\lambda_1^+ + \lambda_2^-),\\
R_{33}&=&-1+2(\lambda_2^+ + \lambda_2^-).\nonumber
\end{eqnarray}

The trace distance discord quantifying quantum correlations of an arbitrary state $\rho_{AB} \equiv \rho$ of a bipartite system $AB$, as revealed on subsystem $A$, can be defined as \cite{noq,sarandy}
\begin{eqnarray}\label{DTD}
D_\mathrm{TD}(\rho)&\equiv& \inf_{\chi \in {\cal C}}\delta_\mathrm{TD}(\rho,\chi) = \delta_\mathrm{TD}(\rho,\chi_\rho) \nonumber \\
&=& \frac{1}{2}\|\rho-\chi_\rho\|_1=
\frac{1}{2}\mathrm{Tr}\sqrt{(\rho-\chi_\rho)^\dag (\rho-\chi_\rho)},
\end{eqnarray}
where ${\cal C}$ the set of classical states $\chi$. By classical states we mean states with zero discord on subsystem $A$, also known as classical-quantum states, which can be written in general as $\chi = \sum_i p_i |i\rangle\langle i |_A \otimes {\tau_i}_B$, with $\{p_i\}$ being a probability distribution,  ${\ket{i}_A}$ an orthonormal basis for subsystem $A$, and ${\tau_i}_B$ an ensemble of arbitrary states for subsystem $B$. We have denoted by $\chi_\rho$ the classical state closest to $\rho$ in trace distance, which achieves the infimum in Eq.~(\ref{DTD}).
Due to the hermiticity of the density matrices, the previous equation is equal to
\begin{equation}\label{TDDiscord}
D_\mathrm{TD}(\rho)=\frac{1}{2}\sum_i |\lambda^D_i|,
\end{equation}
where $\lambda^D_i$ are the eigenvalues of the matrix $(\rho-\chi_\rho)$.

In Ref.~\cite{noq}, it has been proven that when $A$ is a qubit, the trace distance discord is equivalent to the so-called negativity of quantumness, which quantifies the minimum negativity of entanglement \cite{horodecki2009RMP} created with an apparatus during a local projective measurement of subsystem $A$, according to the formalism of \cite{streltsov,acti,sciar}. The same measure also coincides with the minimum trace distance between $\rho$ and the state decohered after a minimally disturbing local measurement,
\begin{equation}\label{NoQprojective}
D_\mathrm{TD}(\rho)=\mathrm{min}_{\Pi_A}\frac{\|\rho-\Pi_A[\rho]\|_1}{2},
\end{equation}
where $\Pi_A$ is a projective measurement on subsystem $A$ \cite{noq}.

A closed expression for the trace distance discord $D_\mathrm{TD}(\rho_{\mathrm{B}})$ for arbitrary Bell diagonal states $\rho_{\mathrm{B}}$ of two qubits was obtained in \cite{noq,sarandy}. One simply has
\begin{equation}\label{TDDiscordfinal}
D_\mathrm{TD}(\rho_\mathrm{B})=\frac{R_\mathrm{int}}{2},
\end{equation}
where $R_\mathrm{int}$ represents the intermediate value among the moduli $|R_{ii}|$ ($i=1,2,3$).

In the following, for completeness, we construct the explicit form of the closest classical state $\chi_{\rho_\mathrm{B}}$ to an arbitrary Bell diagonal state $\rho_\mathrm{B}$, which attains the minimum in Eq.~(\ref{DTD}) resulting in the expression given by  Eq.~(\ref{TDDiscordfinal}) for the trace distance discord.

\subsection{Closest classical state}

It is known in the literature that the classical state closest to $\rho_\mathrm{B}$ according to both the relative entropy distance and the Hilbert-Schmidt distance is still a Bell diagonal state and has the form \cite{bellomo2012PRA,Modi2010PRL}
\begin{equation}\label{classicalstate}
\chi_{\rho_\mathrm{B}}=\left[\mathbb{I}_A\otimes\mathbb{I}_B+R_{kk}\sigma_k\otimes\sigma_k\right]/4,
\end{equation}
where $R_{kk}$ is the one among the elements $R_{11},R_{22},R_{33}$ such that $|R_{kk}|\equiv R_\mathrm{max}=\mathrm{max}\{|R_{11}|,|R_{22}|,|R_{33}|\}$. Notice that the closest classical state $\chi_{\rho_\mathrm{B}}$ above is symmetric under exchange of subsystems $A$, $B$, thus it has vanishing discord when detected either on subsystem $A$ or on subsystem $B$ according to any distance measure \cite{dakic2010}. We now show that this state is also the closest classical state to $\rho_\mathrm{B}$ in the trace distance.

Using the relations of Eq.~(\ref{lambdaR}) among the eigenvalues of a Bell diagonal state and the coefficients $R_{ii}$, one can distinguish different cases in the ordering of the $|R_{ii}|$ and one can correspondingly obtain the expression of the trace distance discord $D_\mathrm{TD}(\rho_\mathrm{B})$ for the class of Bell diagonal states. Let us select indices $i$, $j$ and $k$ as an ordering of $1$, $2$ and $3$ such that $|R_{ii}| \leq |R_{jj}| \leq |R_{kk}|$.
In this case we postulate that the closest classical state assumes the form as in Eq.~(\ref{classicalstate}), from which one gets
\begin{equation}\label{TDdiscord1}
D_\mathrm{TD}(\rho_\mathrm{B})=\frac{1}{4}\left[|R_{jj}-R_{ii}|+|R_{jj}+R_{ii}|\right].
\end{equation}
Notice that flipping the sign of $R_{jj}$ or $R_{ii}$ in the above expression simply swaps the two absolute value terms, thus leaving the entire expression invariant. Therefore, no matter the signs, we obtain for this choice $D_\mathrm{TD}(\rho_\mathrm{B})=\frac{1}{2}R_\mathrm{int}$, where $R_\mathrm{int} = |R_{jj}|$, which matches the expression announced in Eq.~(\ref{TDDiscordfinal}) and computed independently in \cite{noq,sarandy}.
The above calculation shows that the state $\chi_{\rho_\mathrm{B}}$ of Eq.~(\ref{classicalstate}) is indeed the classical state closest to an arbitrary Bell diagonal state $\rho_\mathrm{B}$ in the trace distance. Interestingly, the state of Eq.~(\ref{classicalstate}) is thus the closest classical state to a Bell diagonal state for all the three distances, namely relative entropy, Hilbert-Schmidt and trace distance. In the following section, we see that this similarity among the different metrics is not preserved when classical and total correlations are concerned.

\section{\label{sec:TotClassCorr}Total and classical trace distance correlations}
The definition of quantifiers of total and classical correlations for a bipartite state $\rho$ in geometric terms \cite{Modi2010PRL,bellomo2012PRA2} requires finding the closest product state to $\rho$ and to $\chi_{\rho}$, respectively, where $\chi_{\rho}$ is the classical state closest  to $\rho$ as defined in Eq.~(\ref{DTD}); see Fig.~\ref{fipalla} for a schematic picture. We define by ${\cal P}$ the set of product states $\pi = \gamma_A \otimes \tau_B$, where $\gamma_A$ and $\tau_B$ are arbitrary states defined on the marginal Hilbert spaces of subsystems $A$ and $B$, respectively. Note that ${\cal P} \subset {\cal C} \subset {\cal H}$ in general, where ${\cal H}$ represents the Hilbert space of the composite system $AB$ and ${\cal C}$ contains classical states as defined earlier. Adopting trace distance in the present framework, we can introduce quantifiers of total and classical correlations for a bipartite state $\rho$ as follows
\begin{eqnarray}
T_\mathrm{TD}(\rho)&\equiv & \inf_{\pi \in {\cal P}} \delta_\mathrm{TD}(\rho,\pi) = \delta_\mathrm{TD}(\rho,\pi_\rho)=\frac{1}{2}\|\rho-\pi_{\rho}\|_1 \nonumber \\
&=&\frac{1}{2}\mathrm{Tr}\sqrt{(\rho-\pi_\rho)^\dag (\rho-\pi_\rho)}=\frac{1}{2}\sum_i |\lambda_i^T|\,; \label{TDTotal} \\
& & \nonumber \\
C_\mathrm{TD}(\rho)&\equiv & \inf_{\pi \in {\cal P}} \delta_\mathrm{TD}(\chi_\rho,\pi) = \delta_\mathrm{TD}(\chi_\rho,\pi_{\chi_\rho})=\frac{1}{2}\|\chi_\rho-\pi_{\chi_\rho}\|_1= \nonumber \\
&=&\frac{1}{2}\mathrm{Tr}\sqrt{(\chi_\rho-\pi_{\chi_\rho})^\dag (\chi_\rho-\pi_{\chi_\rho})}=\frac{1}{2}\sum_i |\lambda_i^C|, \label{TDClassical}
\end{eqnarray}
where $\pi_\rho$ and $\pi_{\chi_\rho}$ indicate, respectively, the product state closest to $\rho$ and the product state closest to $\chi_\rho$ in trace distance, while  $\lambda_i^T$ and $\lambda_i^C$ are the eigenvalues of the matrices $(\rho-\pi_\rho)$ and $(\chi_\rho-\pi_{\chi_\rho})$, respectively.

In the next subsections, we derive explicit expressions for Eqs.~(\ref{TDTotal},\ref{TDClassical}) for Bell diagonal states.

\subsection{Classical correlations}
We now find a closed form for $\pi_{\chi_{\rho_{\mathrm{B}}}}$ and the analytical value of $C_\mathrm{TD}$ for Bell diagonal ${\rho_{\mathrm{B}}}$. Defining two arbitrary states of single qubits $A$ and $B$ with corresponding Bloch vectors $\textbf{a}^+ = (a_1, a_2, a_3)$ and $\textbf{b}^+ = (b_1, b_2, b_3)$ as
$\tilde{\rho}_A=\frac{1}{2}[\mathbb{I}_A+\sum_i a_i\sigma_i]$ and
$\tilde{\rho}_B=\frac{1}{2}[\mathbb{I}_B+\sum_i b_i\sigma_i]$,
their product state is
\begin{eqnarray*}\label{arbitraryproductstate}
\!\!&\!\!&\!\!\!\!\pi^+=\tilde{\rho}_A\otimes \tilde{\rho}_B \hfill \\
\!\!&\!\!&\!\!\!\!=\!\!\begin{array}{c}\frac{1}{4}[\mathbb{I}_A\otimes\mathbb{I}_B+\sum_i a_i\sigma_i\otimes \mathbb{I}_B+
\sum_i b_i\mathbb{I}_A\otimes\sigma_{j}+\sum_{i,j}a_i b_j\sigma_{i}\otimes\sigma_{j}]\end{array}\!.
\end{eqnarray*}
For a given product state, we consider a corresponding state $\pi^-$ given by vectors  $\textbf{a}^- = (-a_1, a_2, a_3)$, $\textbf{b}^- = (b_1, b_2, b_3)$. Note that $\frac{1}{2}(\pi^+ + \pi^-)$ is also a product state, $\pi^0$, with vectors  $\textbf{a}^0 = (0, a_2, a_3)$, $\textbf{b}^0 = (b_1, b_2, b_3)$.

We have seen that the state of Eq.~(\ref{classicalstate}) is the closest classical state ${\rho_{\mathrm{B}}}$ to a Bell diagonal state ${\rho_{\mathrm{B}}}$ for the trace norm. By comparison of characteristic polynomials, it can be verified that if $k \neq 1$, then $\chi_{\rho_{\mathrm{B}}} - \pi^+$ has the same eigenvalues as $\chi_{\rho_{\mathrm{B}}} - \pi^-$, where as before $k$ is the index such that $|R_{kk}|\equiv R_\mathrm{max}$. This gives us
\begin{equation}
\|{\chi_{\rho_{\mathrm{B}}}}-\pi^+\|_1=\|{\chi_{\rho_{\mathrm{B}}}}-\pi^-\|_1
\end{equation}
Trace distance also satisfies the convexity property
\begin{equation}
\|A-(\mu B_1 + (1-\mu) B_2)\|_1 \leq \mu \|A-B_1\|_1 + (1-\mu)\|A-B_2\|_1,
\end{equation} with $0 \leq \mu \leq 1$.
Setting $\mu = \frac{1}{2}$, $A=\chi_{\rho_\mathrm{B}}$, $B_1 = \pi^+$, $B_2 = \pi^-$ yields
\begin{equation}
\|{\chi_{\rho_{\mathrm{B}}}} - \pi^0\|_1 \leq \|{\chi_{\rho_{\mathrm{B}}}} - \pi^+\|_1
\end{equation}
Equivalent results can be found when flipping the sign of any other single vector element $a_i$ or $b_j$ for $i,j \neq k$. This means that for the closest product state $\pi_{\chi_\rho}$, only the Bloch vector elements $a_k$ and $b_k$ can be nonzero. Optimizing over these two remaining elements gives the form
\begin{eqnarray} \label{ProdStateForm}
&&a_k = \frac{|R_{kk}|}{R_{kk}} b_k,  \nonumber \\
&&a_i = a_j = b_i = b_j = 0,\,\, i,j \neq k,
\end{eqnarray}
with the specific solution found at
\begin{equation} \label{ProdStateValue}
a_k = \mp 1 \pm \sqrt{1 + R_{\max}}.
\end{equation}
Finally, plugging this state in Eq.~(\ref{TDClassical}) gives
\begin{equation}\label{TDClassicalfinal}
C_\mathrm{TD}(\rho_\mathrm{B}) = -1 + \sqrt{1+ R_{\max}}\,.
\end{equation}
Remarkably, there is a nice division of roles between the intermediate and the maximum correlation matrix element of an arbitrary Bell diagonal state of two qubits: the former entirely characterizes the trace distance discord, while the latter entirely characterizes the trace distance classical correlations.

We notice that the product state $\pi_{\chi_{\rho_\mathrm{B}}}$ closest to the classical state $\chi_{\rho_\mathrm{B}}$ is not the product of its marginals, and is not even a Bell diagonal state in general. This already reveals how minimizing trace distances from the set ${\cal P}$ of product states is a nontrivial problem which can have counterintuitive solutions. This marks a significant difference between the trace distance and the relative entropy and Hilbert-Schmidt distances.

\subsection{Total correlations}

For most metrics, finding the distance between a given composite state and the closest product state is an easier problem compared to, e.g., minimizing the distance from the set of separable or classical states. Adopting the relative entropy, for instance, the distance between a bipartite state $\rho$ and the set of product states returns the mutual information of $\rho$, which is exactly computable, while the relative entropy of entanglement and the relative entropy of discord are generally hard to obtain. It is in this respect quite surprising that the situation is radically different using the trace distance. Notwithstanding its privileged role in quantum statistics \cite{helstrom,ruskai}, it seems that the trace distance does not induce an intuitive characterization of total correlations in bipartite states. In other words, if we are facing the task of distinguishing between a bipartite state $\rho$ and the closest product state in trace distance, the answer is not trivial. In general, the closest product state is not the product of the marginals of $\rho$. This makes the optimization of the distance in Eq.~(\ref{TDTotal}) over ${\cal P}$ complicated already for simple classes of two-qubit states. Here we focus on $\rho$ being an arbitrary Bell diagonal state $\rho_\mathrm{B}$.

\begin{figure}[tb]
\begin{center}
{\includegraphics[width=0.48\textwidth]{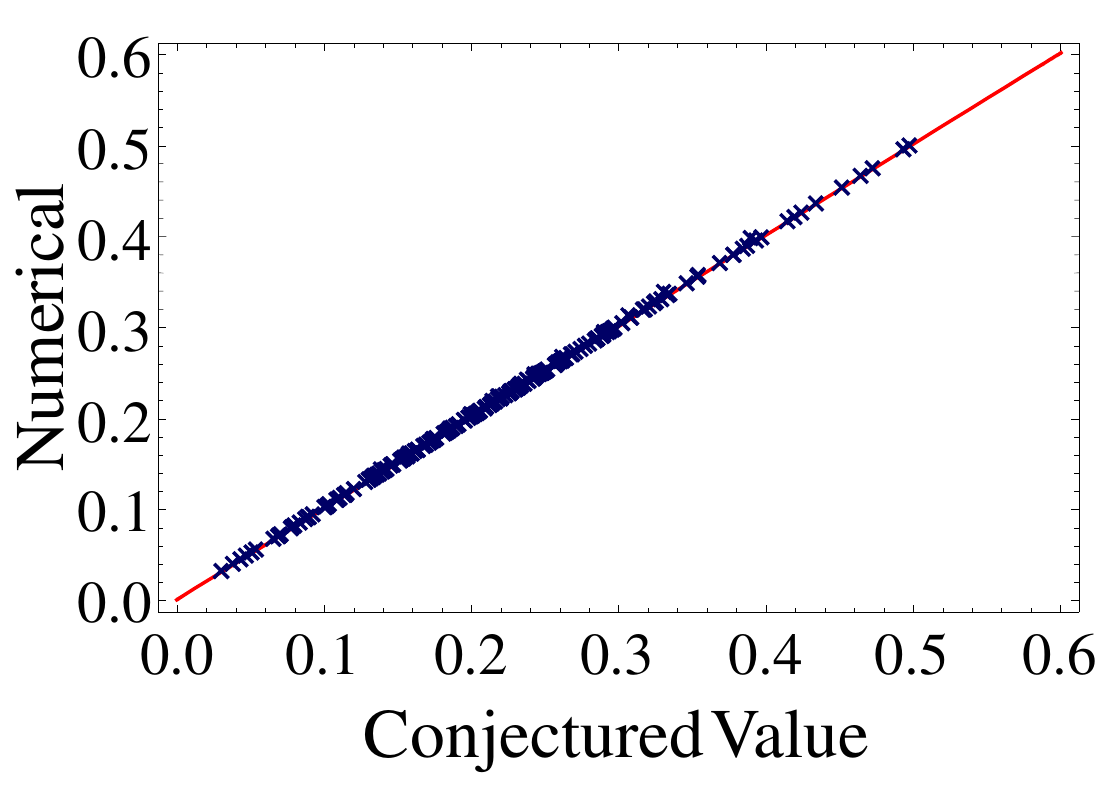}
\caption{\label{fig:TotalCorrelVerif}(Color online) Plot of $T_\mathrm{TD}$ obtained by numerically optimizing the product state closest to a Bell diagonal state $\rho_\mathrm{B}$ against the analytic expression of the trace distance total correlations corresponding to a product state given by Eq.~(\ref{ProdStateForm}) for the same $\rho_\mathrm{B}$. Dark blue crosses show each point (constituting a sample of $10^3$ random  states $\rho_\mathrm{B}$), while a solid red line shows equality between the two.}}
\end{center}
\end{figure}

We base our analytical analysis on an ansatz which is verified numerically. The ansatz is that the closest product state $\pi_{\rho_\mathrm{B}}$ to $\rho_\mathrm{B}$ can be found once more among the states of the form given by  Eq.~(\ref{ProdStateForm}), where as usual $k$ corresponds to $|R_{kk}|\equiv R_\mathrm{max}$. However, while the index of the nonzero Bloch vector element and the relative sign of $a_k$ and $b_k$ depend only on $R_{kk}$, we anticipate that the actual optimal value of $a_k$  determining  $T_\mathrm{TD}$ depends on $R_{ii}$ and $R_{jj}$ as well, which means that in general  $\pi_{\rho_\mathrm{B}} \neq \pi_{\chi_{\rho_\mathrm{B}}}$, as schematically depicted in Fig.~\ref{fipalla}. Under the ansatz of Eq.~(\ref{ProdStateForm}), the total trace distance correlations of a Bell diagonal state $\rho_\mathrm{B}$ can be written as follows
\begin{eqnarray}
\label{TDTotalfinal}
T_\mathrm{TD}(\rho_\mathrm{B}) &=& \min_{a_k: |a_k|\leq 1} \frac{1}{8}\Bigg(\big|a_k^2 + R_{ii} + s(R_{jj} - R_{kk})\big| \\
&+& \big|a_k^2 - R_{ii} + s(-R_{jj} - R_{kk})\big| \nonumber \\&+& \bigg|a_k^2 - s R_{kk} + s \sqrt{4 a_k^2 + (R_{ii} - R_{jj})^2}\bigg| \nonumber
 \\ &+& \bigg|-s a_k^2 + R_{kk} + \sqrt{4 a_k^2 + (R_{ii} - R_{jj})^2}\bigg|\Bigg)\,, \nonumber
\end{eqnarray}
where $s=R_{kk}/|R_{kk}|$ is the sign of $R_{kk}$, and $a_k$ is the product state parameter from Eq.~(\ref{ProdStateForm}). Note that this expression is invariant under interchange of $R_{ii}$ and $R_{jj}$. The remaining optimization in Eq.~(\ref{TDTotalfinal}) can be solved in closed form. It turns out that the the optimum $a_k$ is either $0$ (meaning that the closest product state is the identity, i.e., the product of the marginals of $\rho_\mathrm{B}$), or it has to be found among those values which nullify each of the absolute value terms in Eq.~(\ref{TDTotalfinal}). However, the resulting explicit expression for $T_\mathrm{TD}(\rho_\mathrm{B})$ is too long and cumbersome to be reported here.

It is important to comment on the validity of the ansatz behind Eq.~(\ref{TDTotalfinal}). We have ran an extensive numerical test where we compared the conjectured expression for $T_\mathrm{TD}(\rho_\mathrm{B})$ obtained under the assumption of Eq.~(\ref{ProdStateForm}), with a numerical minimization of Eq.~(\ref{TDTotal}) over arbitrary product states $\pi$ of two qubits. The result for a sample of $10^3$ Bell diagonal states $\rho_\mathrm{B}$ (out of a total of $10^6$ tested ones) is shown in Fig.~\ref{fig:TotalCorrelVerif}: The numerically optimized trace distance for all tested states falls on or above the straight line representing equality with the analytical formula resulting from Eq.~(\ref{TDTotalfinal}), which means that  no product state could be found numerically closer---in trace distance---to a generic Bell diagonal state, than the one analytically given by Eqs.~(\ref{ProdStateForm}, \ref{TDTotalfinal}).

For the majority of states $\rho_\mathrm{B}$, the optimal $a_k = s b_k \neq 0$, which means that the closest product state is {\it not} the product of the marginals, in contrast to the total correlation measures obtained by using the relative entropy or Hilbert-Schmidt norms \cite{Modi2010PRL,bellomo2012PRA}.
 Additionally, the triangle inequality for trace distance implies in general
  \begin{equation}
  \label{ineq}
  T_\mathrm{TD}(\rho_\mathrm{B}) \leq C_\mathrm{TD}(\rho_\mathrm{B}) + D_\mathrm{TD}(\rho_\mathrm{B})\,,
  \end{equation}
but unlike the other norms the inequality is typically sharp for the trace distance case. In this respect, we wish to point out that this is not a byproduct of the ansatz used to derive Eq.~(\ref{TDTotalfinal}). Even though a closer product state to $\rho_\mathrm{B}$ might be found (which appears extremely unlikely based on our numerical analysis), the expression in Eq.~(\ref{TDTotalfinal}) would remain an upper bound to the true trace distance total correlations, therefore not altering the sharpness of the inequality (\ref{ineq}).

Hereby we will confidently regard  the value of $T_\mathrm{TD}(\rho_\mathrm{B})$ given by Eq.~(\ref{TDTotalfinal}) as the exact value of the trace distance total correlations for arbitrary Bell diagonal states $\rho_\mathrm{B}$.

\subsection{Examples}

Here we present some explicit examples where we compute quantum, classical and total correlations in particular families of two-qubit Bell diagonal states and comment on their properties.

One simple class of Bell diagonal states is Werner states \cite{werner1989PRA}, for which $R_{11} = -R_{22} = R_{33} = r$, for $0 \leq r \leq1$. For these states, the values of the correlations are:
\begin{eqnarray}
D_\mathrm{TD}(\rho) &=& \frac{r}{2};\nonumber \\
C_\mathrm{TD}(\rho) &=& \sqrt{1+r}-1;\\
T_\mathrm{TD}(\rho) &=& \begin{cases}
\frac{3}{4}r, & 0 \leq  r \leq \frac{4}{5}; \\
\frac{1}{2}\sqrt{r+r^2}, & \frac{4}{5} \leq r \leq 1. \nonumber
\end{cases}
\end{eqnarray}

\begin{figure}[tb]
\begin{center}
{\includegraphics[width=0.48\textwidth]{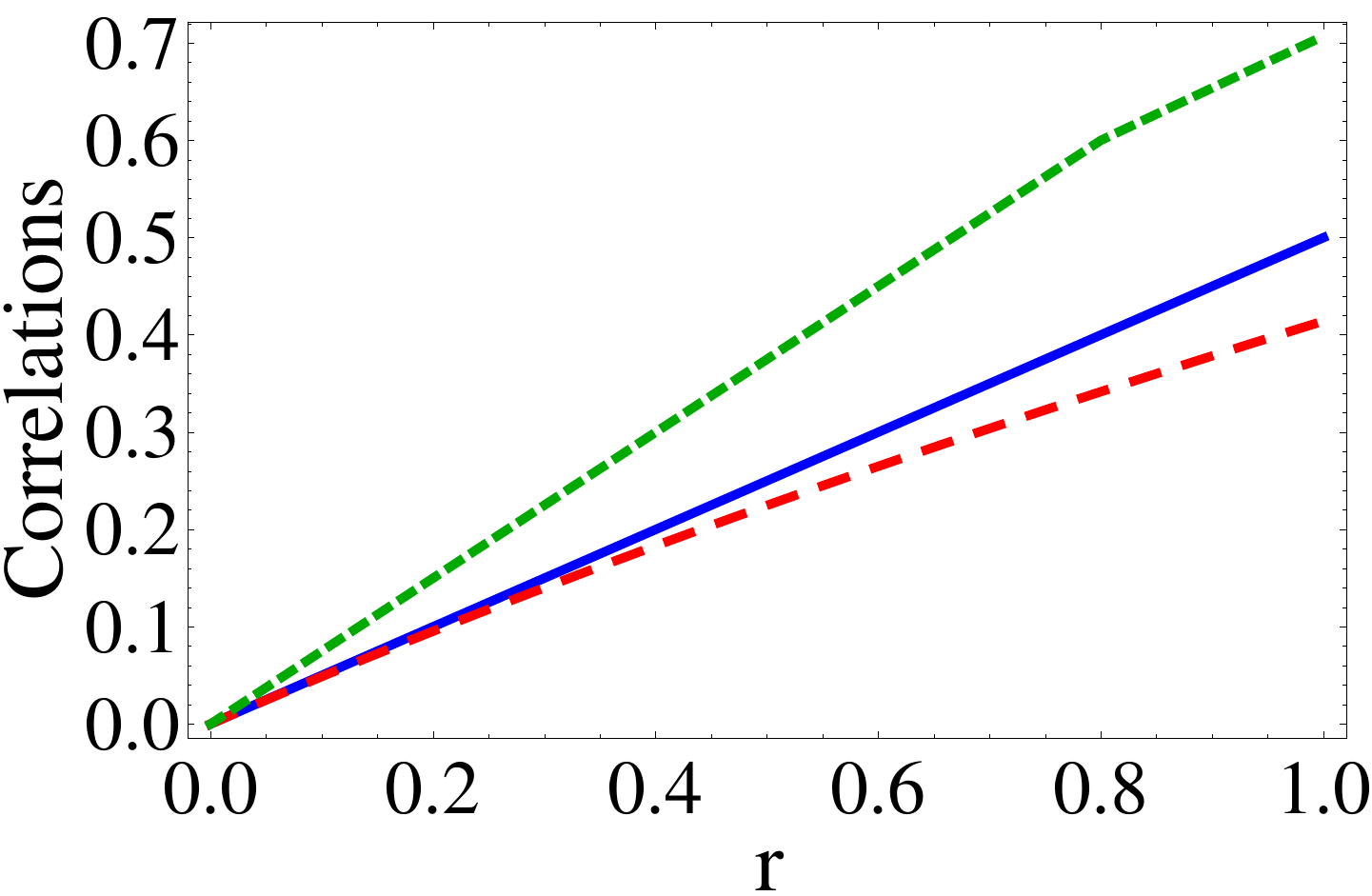}\hspace*{0.04\textwidth}
\includegraphics[width=0.48\textwidth]{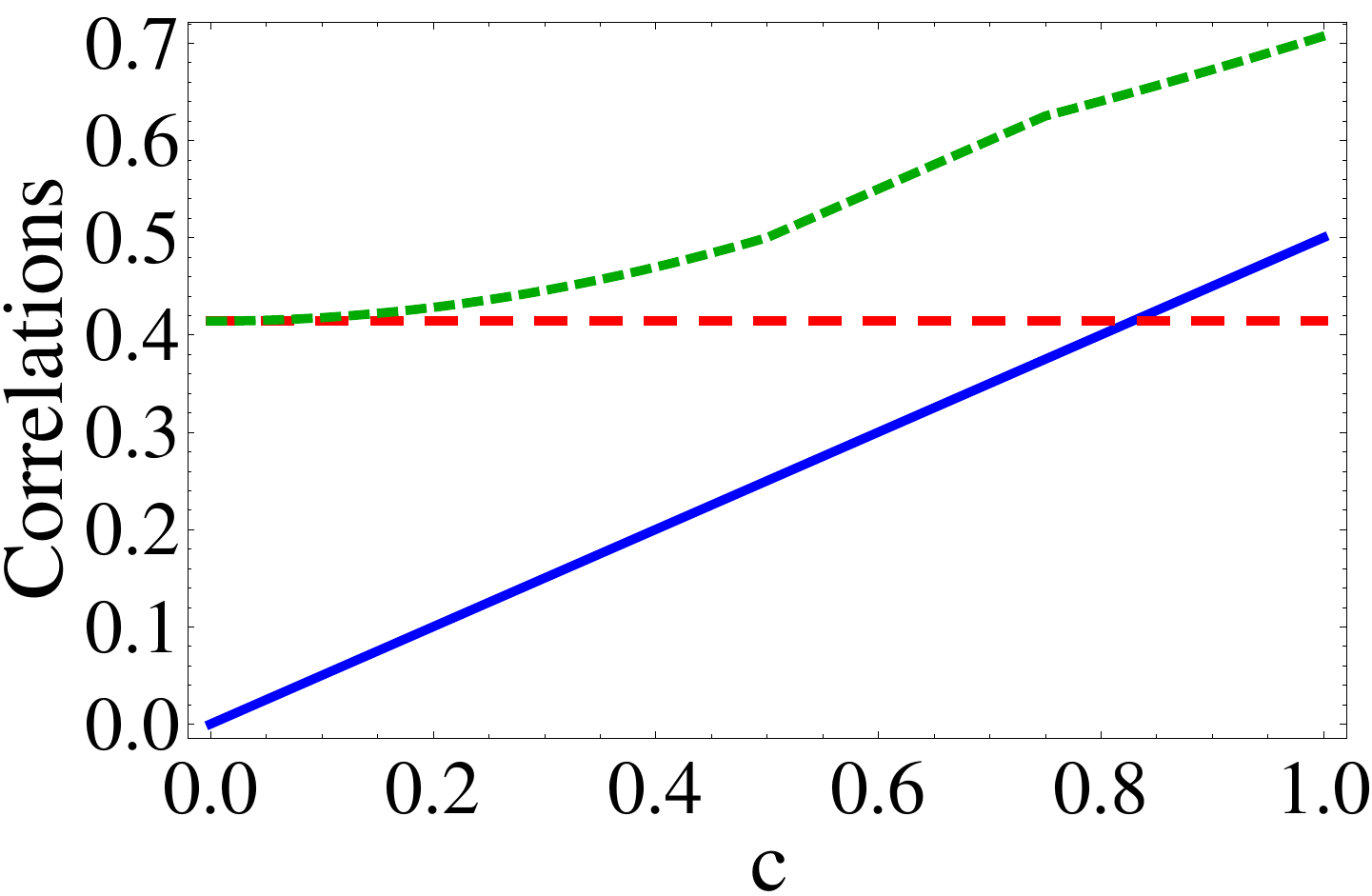}}
\caption{\label{fig:Rank2andWerner}(Color online)  Plot of $D_\mathrm{TD}$ (solid blue line), ${C}_\mathrm{TD}$ (dashed red line) and  ${T}_\mathrm{TD}$ (dotted green line) for Werner and rank-$2$ Bell diagonal states.  Left panel: Werner states with $R_{11} = -R_{22} = R_{33} = r$.  Right panel: Rank-2 Bell diagonal states with $R_{11} = -R_{22} = c, R_{33}=1$}
\end{center}
\end{figure}

This is shown in Fig.~\ref{fig:Rank2andWerner}~(Left). It is notable that, while quantum and classical correlations increase smoothly, the total correlations have a sudden change point at $r=\frac{4}{5}$. This point marks the transition from the region for which the closest product state is the product of the marginals, $0\leq r\leq \frac{4}{5}$, to the region where it is instead a product state of the form as in Eq.~(\ref{ProdStateForm}) with $a_k = \sqrt{r}$.

A second simple class of states, also displayed in Fig.~\ref{fig:Rank2andWerner}~(Right), are the rank-$2$ Bell diagonal states, for which $R_{11} = -R_{22} = c, R_{33} = 1$, for $0 \leq c \leq1$. The values of trace distance correlations for these are
\begin{eqnarray}
D_\mathrm{TD}(\rho) &=& \frac{c}{2};\nonumber \\
C_\mathrm{TD}(\rho) &=& \sqrt{2} - 1; \\
T_\mathrm{TD}(\rho) &=& \begin{cases}
\sqrt{2+c^2}-1, & 0 \leq c \leq \frac{1}{2};\\
\frac{1}{4}\big(1 + 2c \big), & \frac{1}{2} \leq c \leq \frac{3}{4}; \\
\frac{1}{2}\sqrt{1+c^2}, & \frac{3}{4} \leq c \leq 1.
\end{cases} \nonumber
\end{eqnarray}

Here we see that while again quantum correlations increase smoothly, for total correlations there are actually three regions, with two sudden changes. In this case it is the middle region, with $\frac{1}{2} \leq c \leq \frac{3}{4}$, for which the closest product state is the product of the marginals. In the final region, $\frac{3}{4} \leq c \leq 1$, the closest product state is pure. We can additionally note that for both Werner and rank-$2$ Bell diagonal states, it is always the case that  $T_\mathrm{TD} \neq D_\mathrm{TD} + C_\mathrm{TD}$ except for the trivial cases where  one of them vanishes. Notably, classical correlations are constant for rank-$2$ Bell diagonal states.

\section{\label{sec:DynamQCorr}Dynamics of trace distance quantifiers of correlations}
In this section we analyze the dynamics of the trace distance quantifiers of correlations in two specific models exhibiting non-Markovian evolutions and compare them to the dynamics of the correlations measured by relative entropy $S(\rho\|\sigma)\equiv-\mathrm{Tr}(\rho\log \sigma)-S(\rho)$ \cite{Modi2010PRL}, where $S(\rho)\equiv-\mathrm{Tr}(\rho\log\rho)$ is the von Neumann entropy.

This analysis will serve the purpose to highlight possible peculiarities in the dynamical behaviors of the trace distance quantifiers of correlations and to show possible qualitative differences with the dynamics of the entropic ones. The choice of the entropic quantifiers of correlations for the dynamical comparison is due to the fact that both relative entropy and trace distance measures are contractive for any trace-preserving completely positive map $\Lambda$, that is $S(\Lambda \rho\|\Lambda\sigma)\leq S(\rho\|\sigma)$, $\delta_\mathrm{TD}(\Lambda\rho,\Lambda\sigma)\leq\delta_\mathrm{TD}(\rho,\sigma)$: that is a required property for any \emph{bona fide} distance-based measure of correlations \cite{aaronsonPRA}. For instance this property is not exhibited by the Hilbert-Schmidt distance, used to define the geometric discord \cite{dakic2010} which was as such revealed to be an unsuitable measure of quantum correlations \cite{piani2012,tufodiscord}. It is also worth to mention that the relative entropy is adopted as a measure of distance between two states $\rho$, $\sigma$ even if it is asymmetric with respect to the exchange $\rho\leftrightarrow\sigma$ and is thus a pseudo-distance: moreover, $S(\rho\|\sigma)$ diverges when $\sigma$ is a pure state \cite{Eisert2003JPA}. Differently, the trace distance is symmetric to the exchange $\rho\leftrightarrow\sigma$ and it does not present singularities when $\sigma$ or $\rho$ are pure states.

Total correlations $T$, discord $D$ and classical correlations $C$ based on relative entropy are defined as \cite{Modi2010PRL}
\begin{eqnarray}\label{entropicquantifiers}
D(\rho)&\equiv& S(\rho\|\tilde{\chi}_\rho)=S(\tilde{\chi}_\rho)-S(\rho),\nonumber\\
C(\rho)&\equiv& S(\tilde{\chi}_\rho\|\tilde{\pi}_{\tilde{\chi}_\rho})=S(\tilde{\pi}_{\tilde{\chi}_\rho})-S(\tilde{\chi}_\rho),\\
T(\rho)&\equiv& S(\rho\|\tilde{\pi}_\rho)=S(\tilde{\pi}_\rho)-S(\rho),\nonumber
\end{eqnarray}
where $\tilde{\pi}_\rho \in {\cal P}$, $\tilde{\chi}_\rho \in {\cal C}$ are, respectively, the product state and the classical state closest to $\rho$, while $\tilde{\pi}_{\tilde{\chi}_\rho} \in {\cal P}$ is the product state closest to $\tilde{\chi}_\rho$. These states are such that they minimize the corresponding relative entropies, and do not in general coincide with the ones minimizing the trace distance measures of correlations in Eqs.~(\ref{DTD}, \ref{TDTotal}, \ref{TDClassical}). It is worth to notice here that, for the class of Bell diagonal states $\rho_\mathrm{B}$, $D(\rho_\mathrm{B})$ coincides with the original definition of quantum discord \cite{Zurek2001PRL,Henderson2001JPA} and the relative entropy correlation quantifiers satisfy the additivity relation: $T=D+C$ (an analogous relation also holds when using geometric quantifiers defined via the Hilbert-Schmidt distance \cite{bellomo2012PRA}).
The explicit expressions of the entropic correlation quantifiers for Bell diagonal states are \cite{Luo2008PRA}
\begin{eqnarray}\label{BDentropicquantifiers}
D(\rho_\mathrm{B})&=&T(\rho_\mathrm{B})-C(\rho_\mathrm{B}),\nonumber\\
C(\rho_\mathrm{B})&=&\sum_{i=1}^2 \frac{1+(-1)^i R_\mathrm{max}}{2}\mathrm{log}[1+(-1)^i R_\mathrm{max}], \nonumber\\
T(\rho_\mathrm{B})&=&2+\sum_{j,r}\lambda_j^r\log_2\lambda_j^r, \quad (j=1,2; r=\pm)
\end{eqnarray}
where $\lambda_j^r$  and $R_\mathrm{max}$ are defined, respectively, in Eq.~(\ref{lambdaR}) and after Eq.~(\ref{classicalstate}). We shall take into account two different models, a dynamics under local phase-flip channels and an environment of random external fields.

\subsection{First model: phase-flip channels}
We take two noninteracting qubits under local identical phase-flip channels \cite{bellomo2012PRA}. Phase-flip noise, i.e., pure dephasing, is an emblematic type of nondissipative decoherence \cite{nielsenchuang} which arises naturally in typical solid state implementations, such as the case of superconducting qubits interacting with impurities under random telegraph noise \cite{mazzola2010frozen}. In our setting, each qubit is subject to a time-dependent phenomenological Hamiltonian \cite{daffer} $H(t)=\hbar\Gamma(t)\sigma_z$, where $\sigma_z$ is a Pauli operator and $\Gamma(t)=\alpha n(t)$ where $\alpha$ is a coin-flip random variable taking the values $\pm |\alpha|$ while $n(t)$ is a random variable having a Poisson distribution with mean value equal to the dimensionless time $\nu=t/2\tau$. This two-qubit system is characterized by a non-Markovian dynamics that maintains the system inside the class of Bell-diagonal states with the three coefficients $R_{ii}(t)$ of Eq.~(\ref{BelldiagonalstateBloch}) evolving as
\begin{equation}\label{timedependentR}
 R_{i'i'}(t)=R_{i'i'}(0)f^2(\nu), \quad R_{33}(t)=R_{33}(0),
\end{equation}
where $i'=1,2$ and $f(\nu)=\mathrm{e}^{-\nu}[\cos(\mu \nu)+\sin(\mu \nu)/\mu]$ with $\mu=\sqrt{(4\alpha\tau)^2-1}$. Using Eqs.~(\ref{timedependentR}) and (\ref{BDentropicquantifiers}) the quantum correlations can be analytically computed. We notice that the closest classical state $\chi_{\rho_\mathrm{B}(t)}$ of Eq.~(\ref{classicalstate}) is frozen during the time intervals when $|R_{33}(t)|>R_\mathrm{max}$, being $R_{kk}(t)=R_{33}(t)=R_{33}(0)$.

\begin{figure}[tb]
\begin{center}
{\includegraphics[width=0.48\textwidth]{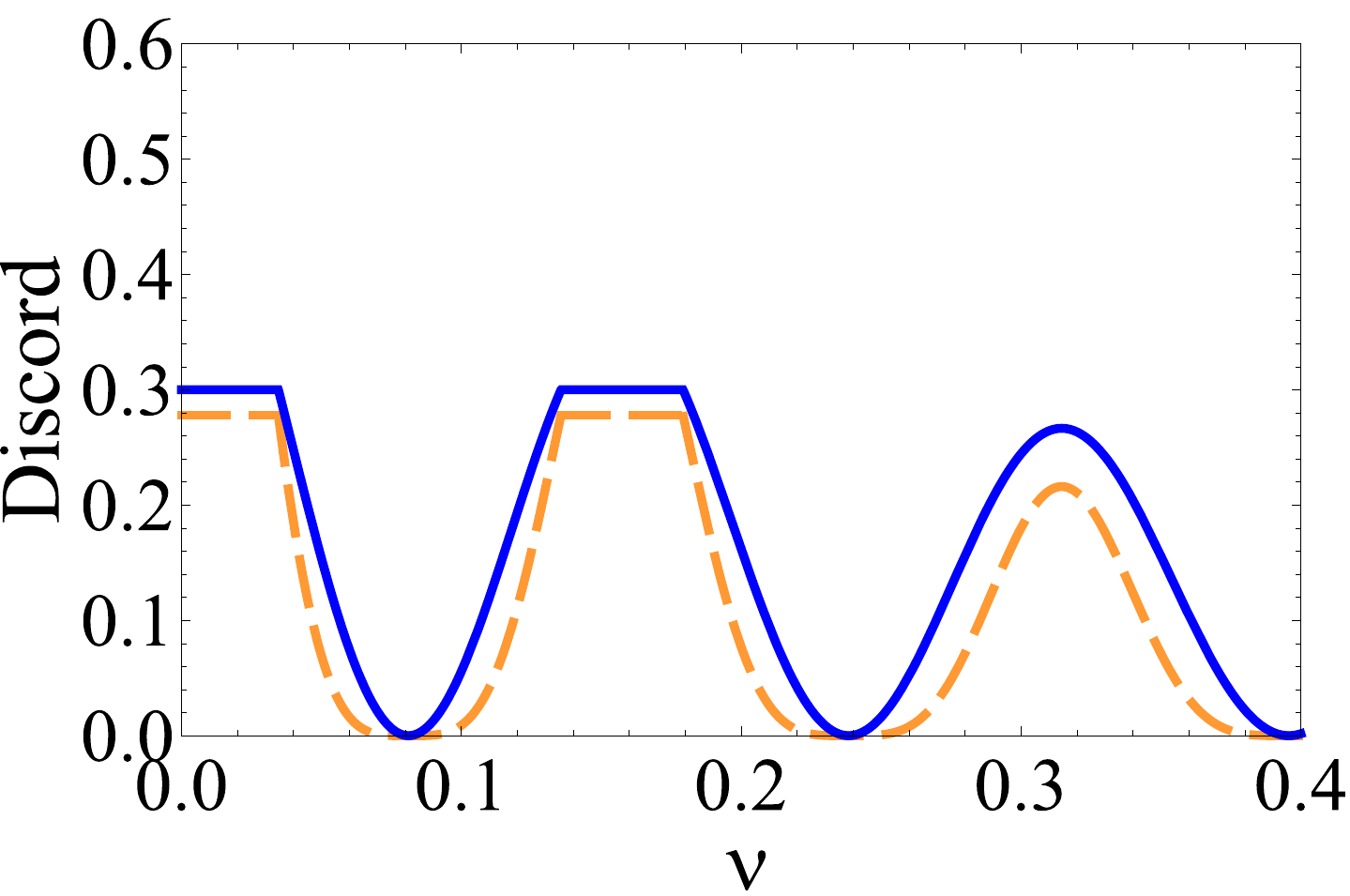}\hspace*{0.04\textwidth}
\includegraphics[width=0.48\textwidth]{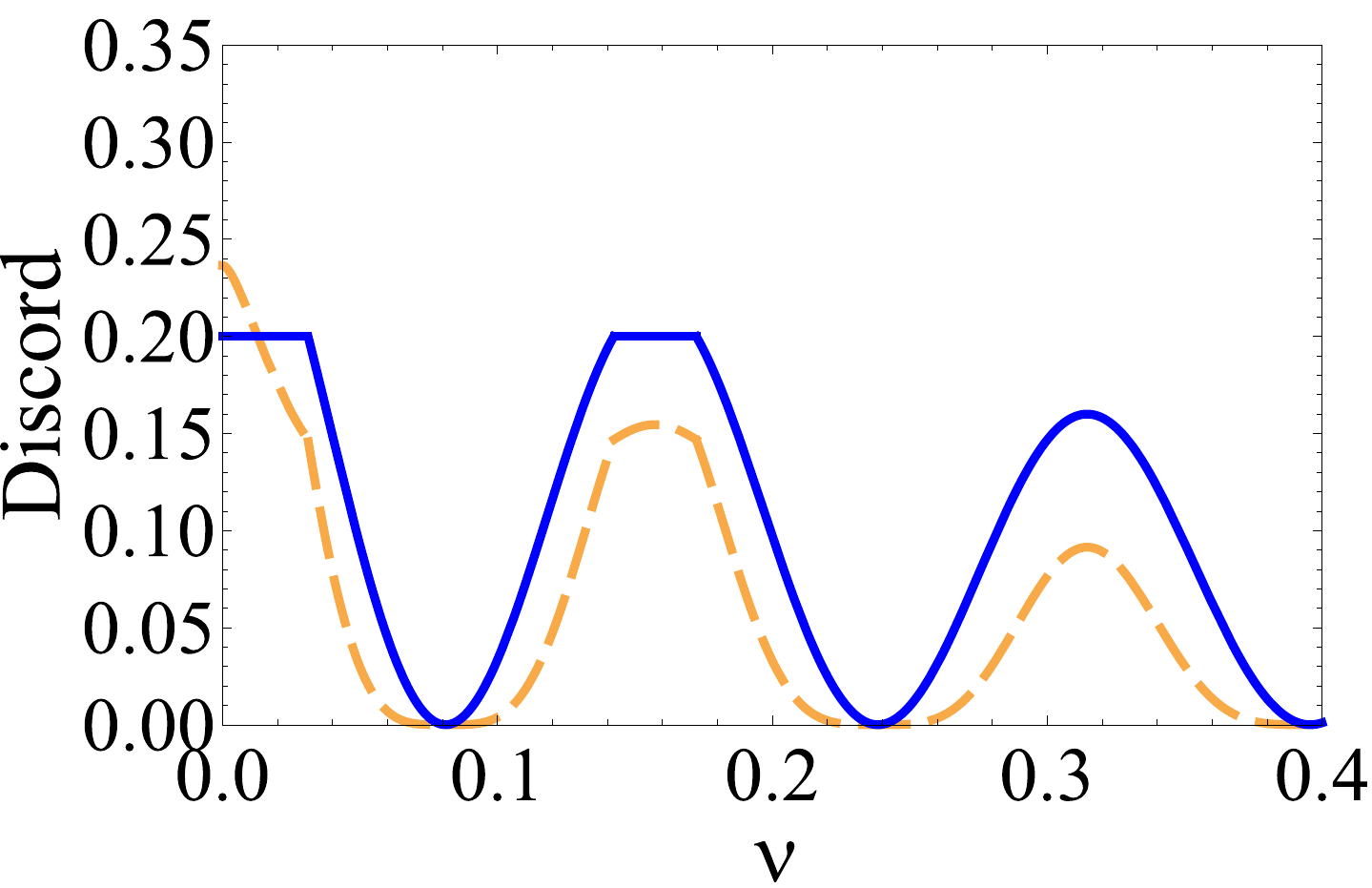}}
\caption{\label{fig:D12}(Color online) Dynamics of discord $D_\mathrm{TD}$ (blue solid line) and $D$ (orange dashed line) under local phase-flip channels versus $\nu=t/2\tau$, with $\tau=5$s and $|\alpha|=1$s$^{-1}$. Left panel: The initial coefficients of the Bell diagonal state are $R_{11}(0)=1$, $R_{22}(0)=-0.6 $ and $R_{33}(0)=0.6$ ($\lambda_1^+(0)=0.2$, $\lambda_1^-(0)=\lambda_2^-(0)=0$, $\lambda_2^+(0)=0.8$).
Right panel: The initial coefficients are $R_{11}(0)=0.6$, $R_{22}(0)=0$ and $R_{33}(0)=0.4$ ($\lambda_1^+(0)=0.3$, $\lambda_1^-(0)=0$, $\lambda_2^+(0)=0.5$, $\lambda_2^-(0)=0.2$).}
\end{center}
\end{figure}

In Fig.~\ref{fig:D12} entropic discord $D$ and trace distance discord $D_\mathrm{TD}$ are plotted as a function of the dimensionless time $\nu$ for two different initial Bell diagonal states.
It is displayed that the entropic discord $D$ assumes different behaviors when the initial conditions are changed (see the orange dashed lines of the two panels), while the trace-norm discord $D_\mathrm{TD}$ maintains the time regions when it is constant.  It is straightforward to see that the initial coefficients $R_{ii}(0)$ of the Bell diagonal state in the Left panel, where the two discord quantifiers are constant in the same time regions, satisfy the condition of freezing for all the \emph{bona fide} quantifiers of quantum correlations under nondissipative evolutions, as introduced in Ref.~\cite{aaronsonPRA} generalizing the seminal analysis of Ref.~\cite{mazzola2010PRL}. Notice also that the two discords display different qualitative behaviors in the Right panel of Fig.~\ref{fig:D12}: while $D_\mathrm{TD}$ is constant, $D$ has sudden changes but no freezing regions.   This demonstrates that the freezing property occurs for a wider range of initial conditions for trace distance discord than for entropic discord. This phenomenon has also been pointed out in Ref.~\cite{sarandydec}, where a richer phenomenology of trace distance discord compared to other measures of discord was uncovered, including the possibility of double sudden changes when phase-flip is combined with amplitude damping. In general, our recent geometric analysis in \cite{aaronsonPRA} shows that, within the space of Bell diagonal states, the trace distance discord $D_\mathrm{TD}$ has broader subregions in which it remains constant compared to any other {\it bona fide} measure of discord. This clearly results in the possibility of larger freezing intervals under various dynamical trajectories compared to other measures. We will now investigate whether this is the case for the second dynamical model studied in this work.

\subsection{Second model: random external fields}
We consider a pair of noninteracting qubits each locally coupled to a random external field, whose characteristics are unaffected by the qubit it is coupled to. This implies that back-action on the dynamics of the qubits is absent \cite{lofranco2013review,lofranco2012PRA}. Each environment is a classical field mode with amplitude fixed and equal for both qubits. The phase of each mode is not determined, and is equal either to zero or to $\pi$ with probability $p=1/2$. This model describes a special case of two qubits each subject to a phase noisy laser \cite{bellomo2012PhysScripErika} but where the phase can take only two values and with the diffusion coefficient in the master equation equal to zero. It has been considered to study revivals of entanglement without back-action \cite{modirev,lofranco2013review,lofranco2012PRA}.

In this model, the dynamical map for the single qubit $S=A,B$ is of the random external fields type \cite{alickibook} and can be written as
\begin{equation}
\label{eq:singlemap}
\Lambda^S_t\rho_S(0)=\frac{1}{2}\sum_{i=1}^2U_i^{S}(t)\rho_S(0)U_i^{S\dag}(t),
\end{equation}
where $U_i^{S}(t)=\mathrm{e}^{-\mathrm{i}H_it/\hbar}$ is the time evolution operator, with $H_i=\mathrm{i}\hbar g(\sigma_+e^{-\mathrm{i}\phi_i}-\sigma_-e^{\mathrm{i}\phi_i})$, and the factor $1/2$ arises from the equal field phase probabilities (there is a probability $p_i^S=1/2$ associated to each $U_i^S$). Each Hamiltonian $H_i$ is expressed in the rotating frame at the qubit-field resonant frequency $\omega$. In the basis $\{\ket{1},\ket{0}\}$, the time evolution operator $U_i^{S}(t)$ has the matrix form
\begin{equation}\label{unitarymatrix}
U_i^{S}(t)=\left(
\begin{array}{cc}\cos(gt)&\mathrm{e}^{-\mathrm{i}\phi_i}\sin(gt)\\
-\mathrm{e}^{\mathrm{i}\phi_i}\sin(gt) & \cos(gt) \\\end{array}\right),
\end{equation}
where $i=1,2$ with $\phi_1=0$ and $\phi_2=\pi$. The single-qubit map $\Lambda_t^S$ generates a nondissipative non-Markovian evolution described by a master equation in a generalized Lindblad form \cite{mannone2013}.
The overall dynamical map $\Lambda_t$ applied to an initial state $\rho(0)$ of the two-qubit system, $\rho(t)\equiv\Lambda_t\rho(0)$, is composed by the two local maps $\Lambda_t^S$ and reads
\begin{equation}\label{globalrandomfieldmap}
\rho(t)=\frac{1}{4}\sum_{i,j=1}^2U_i^{A}(t)U_j^{B}(t)\rho(0)U_i^{A\dag}(t)U_j^{B\dag}(t).
\end{equation}
This map moves inside the class of Bell diagonal states \cite{lofranco2012PRA}. The three coefficients $R_{ii}(t)$ of Eq.(\ref{BelldiagonalstateBloch}) evolve as
\begin{equation}\label{Rtime}
R_{jj}(t)=R_{jj}(0)\cos^2(2gt),\quad R_{22}(t)=R_{22}(0),
\end{equation}
where $j=1,3$.

\begin{figure}[tb]
\begin{center}
{\includegraphics[width=0.48\textwidth]{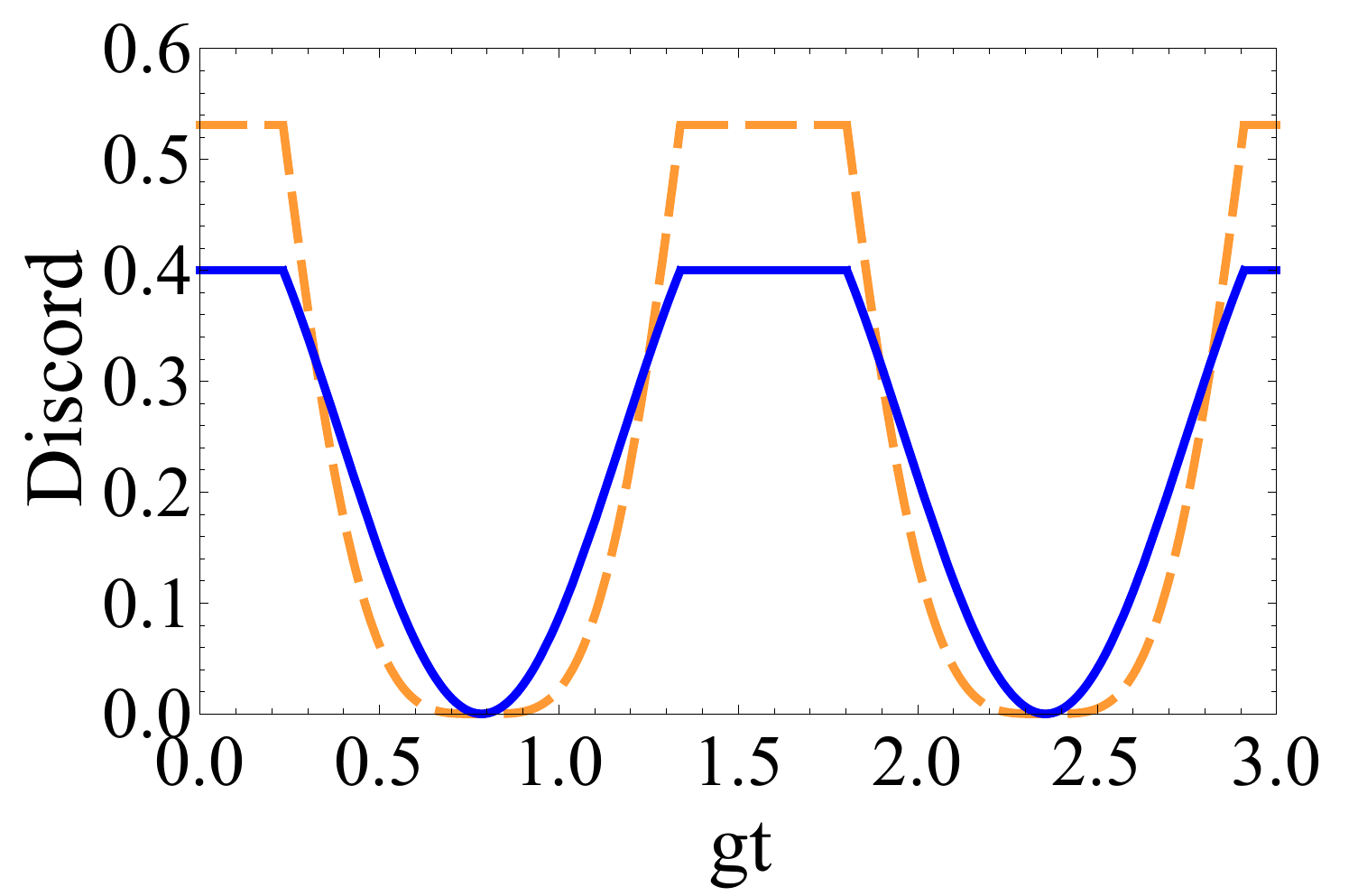}\hspace*{0.04\textwidth}
\includegraphics[width=0.48\textwidth]{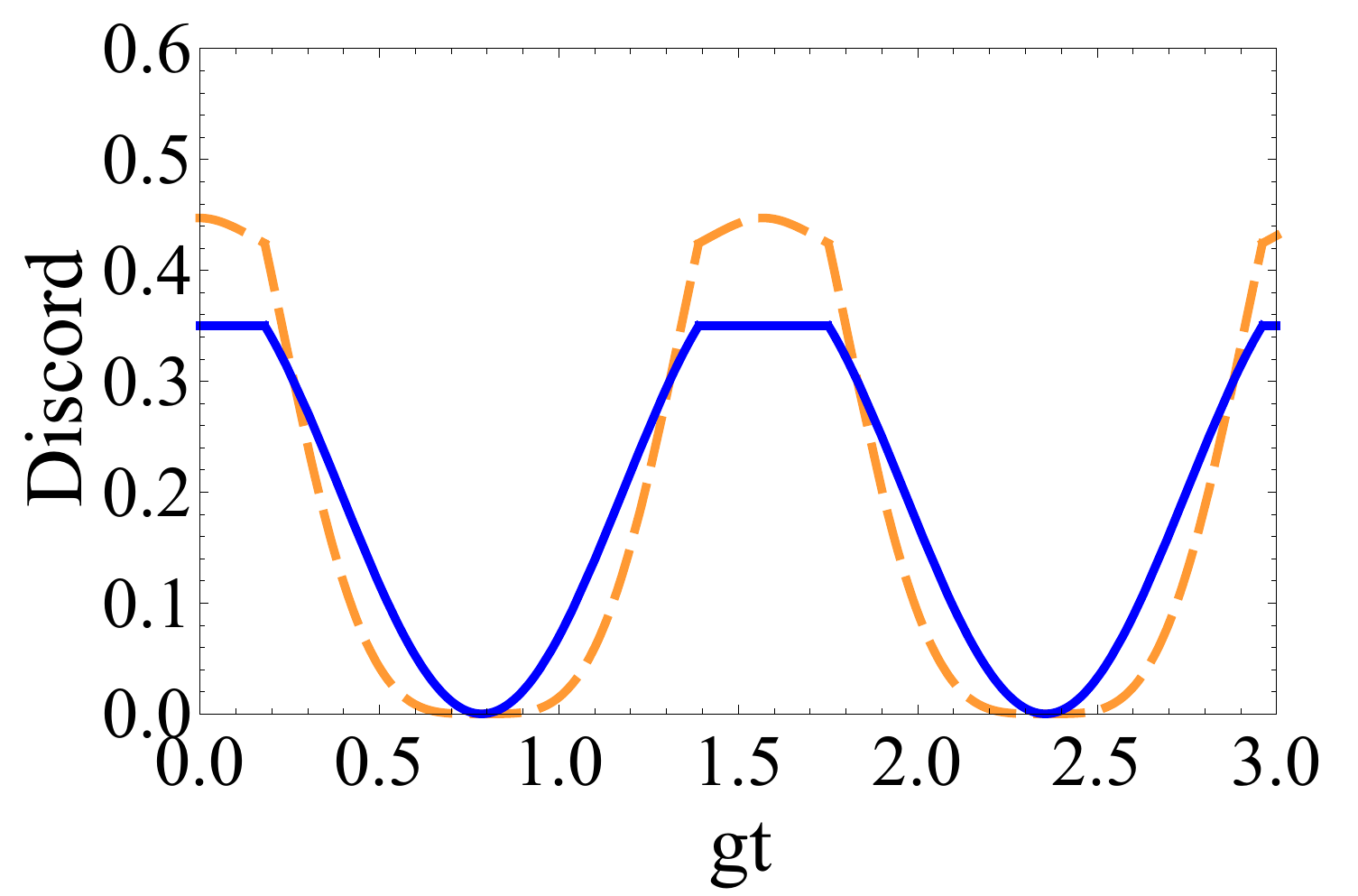}}
\caption{\label{fig:DREF}(Color online) Left panel: Dynamics of discord $D_\mathrm{TD}$ (blue solid line) and $D$ (orange dashed line)  for an initial Bell diagonal state with $\lambda_1^+(0)=0.9$, $\lambda_1^-(0)=0.1$ and $\lambda_2^\pm(0)=0$ (that is, $R_{11}(0)=R_{22}(0)=0.8$, $R_{33}(0)=-1$). Right panel: Dynamics of discord $D_\mathrm{TD}$ (blue solid line) and $D$ (orange dashed line)  for an initial Bell diagonal state with $\lambda_1^+(0)=0.1$, $\lambda_1^-(0)=0.8$ and $\lambda_2^\pm(0)=0.05$ (that is, $R_{11}(0)=R_{22}(0)=-0.7$, $R_{33}(0)=-0.8$).}
\end{center}
\end{figure}

In Fig.~\ref{fig:DREF} we plot entropic discord and trace distance discord for two different initial conditions. Even in this case, as occurred in the above phase-flip channels, the entropic discord changes its qualitative time behavior for the two different initial conditions (time regions of freezing in the Left panel, increase and decrease in the Right panel), while the trace distance discord maintains the same qualitative dynamics. This is a further confirmation of the  fact that the freezing property for trace distance discord occurs for a wider range of initial conditions than for entropic discord.

Once again, it is possible to show that the freezing for both $D$ and $D_\mathrm{TD}$ (Left panel of Fig.~\ref{fig:DREF}) occurs when the initial coefficients $R_{ii}(0)$ of the Bell diagonal state satisfy the general condition of freezing for quantum correlations under nondissipative evolutions \cite{aaronsonPRA}. Notice that the sudden changes in the slope of the two discords occur at the same times; these times can be analytically found for given initial conditions \cite{mazzola2010PRL,aaronsonPRA}.

\subsection{\label{sec:TotClassDynam}Dynamics of total and classical correlations measured by trace distance}

\begin{figure}[tb]
\begin{center}
{\includegraphics[width=0.48\textwidth]{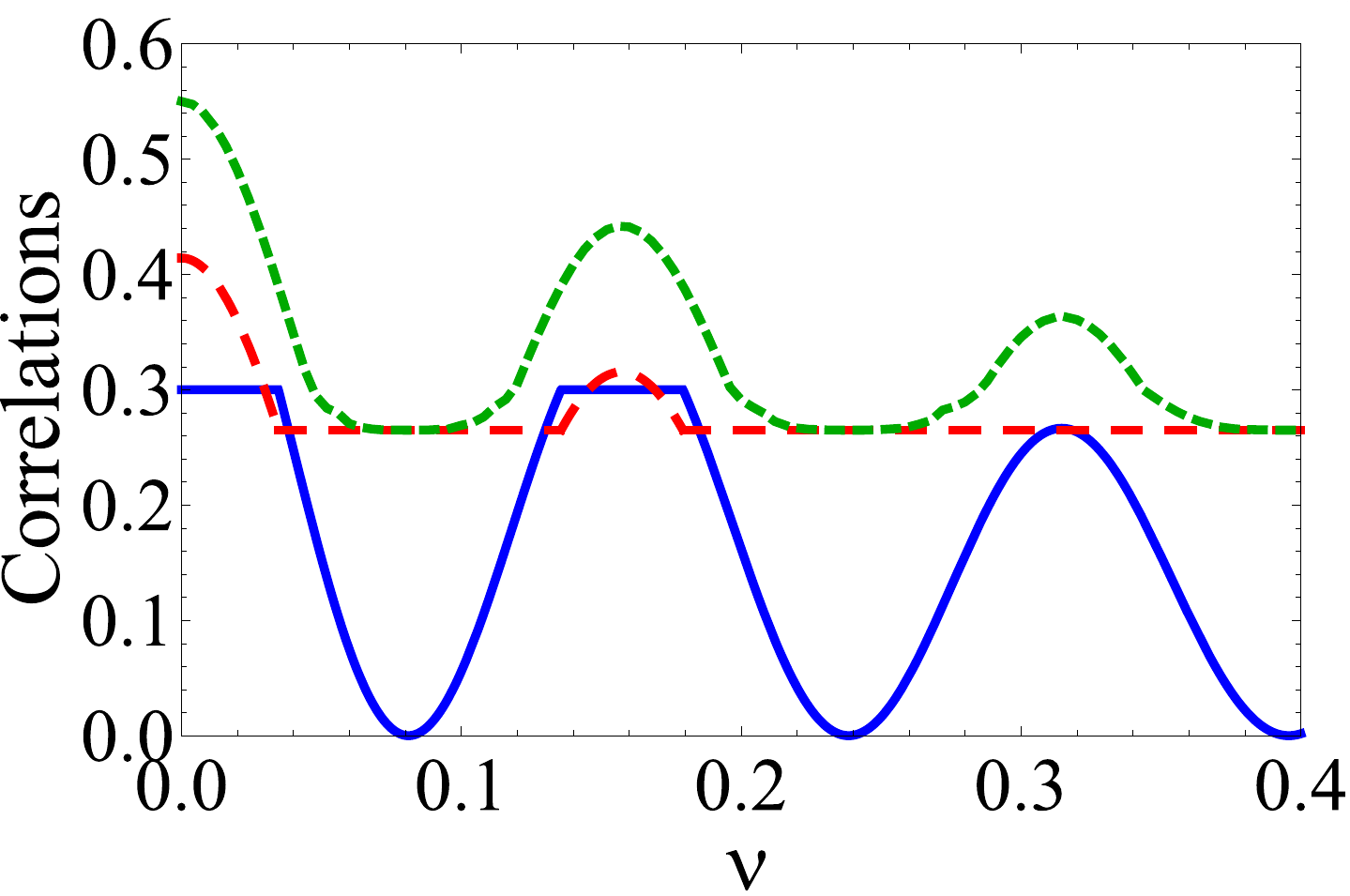}\hspace*{0.04\textwidth}
\includegraphics[width=0.48\textwidth]{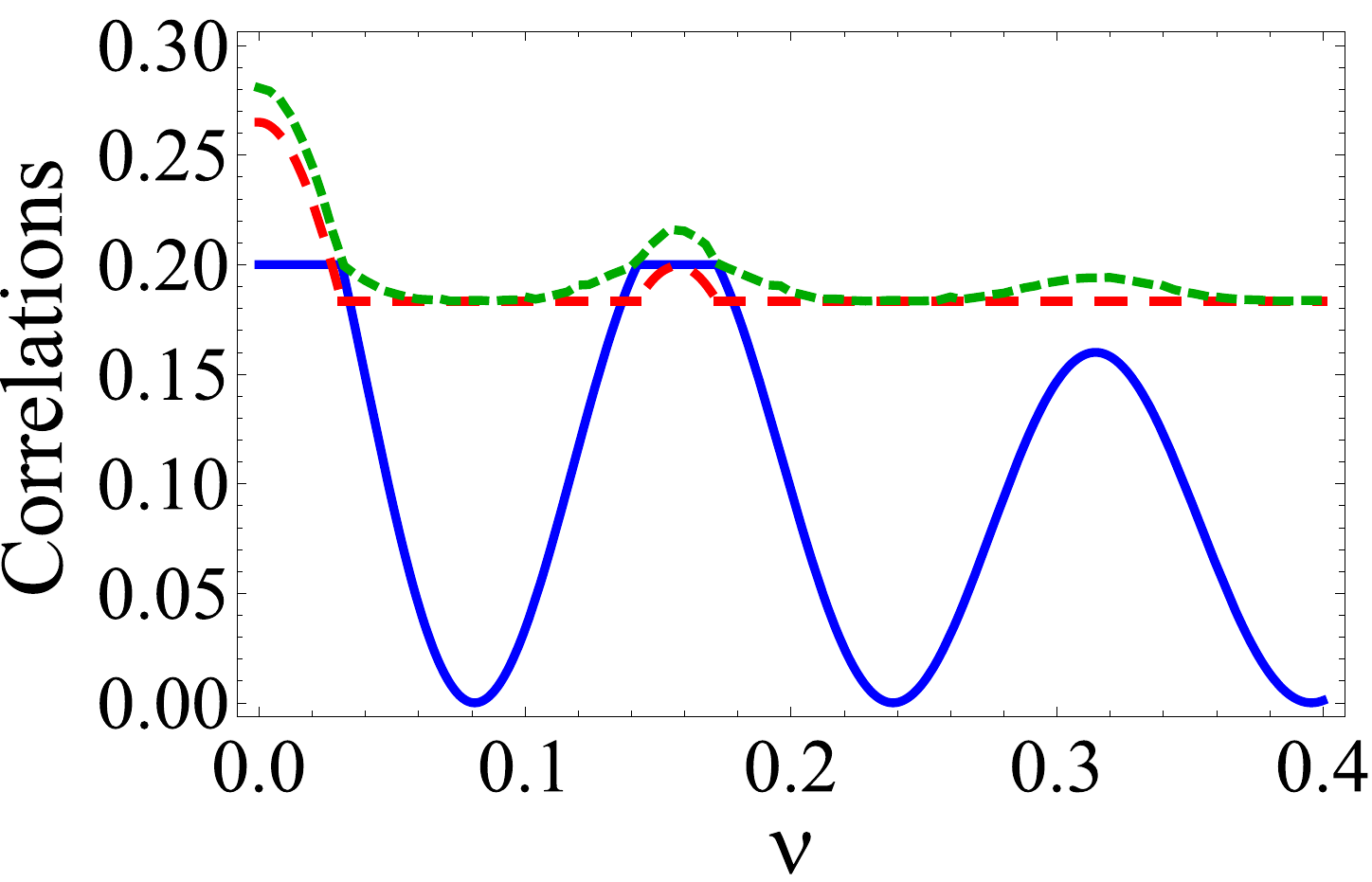}}
\caption{\label{fig:Corr12}(Color online) Plot of  $D_\mathrm{TD}$ (solid blue line), $C_\mathrm{TD}$ (dashed red line) and  $T_\mathrm{TD}$ (dotted green line) for the model of local phase-flip channels, with $\tau=5$s and
$|\alpha|=1$s$^{-1}$. Left panel: Same conditions as in the Left panel of Fig.~\ref{fig:D12}. Right panel: Same conditions as in the Right panel of Fig.~\ref{fig:D12}.}
\end{center}
\end{figure}

\begin{figure}[tb]
\begin{center}
{\includegraphics[width=0.48\textwidth]{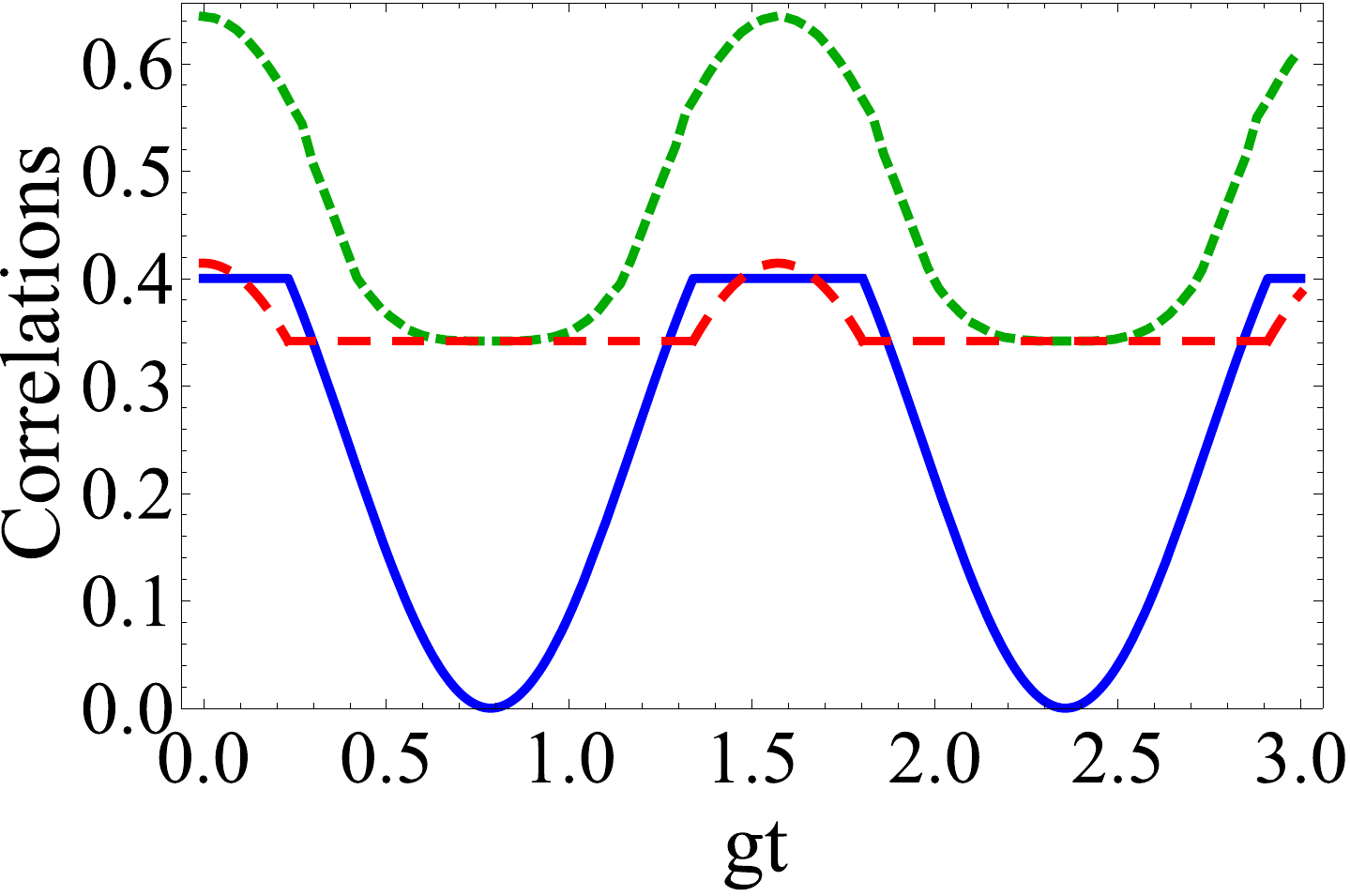}}
\caption{\label{fig:CorrREF}(Color online) Plot of  $D_\mathrm{TD}$ (solid blue line), $C_\mathrm{TD}$ (dashed red line) and  $T_\mathrm{TD}$ (dotted green line) for the model of random external fields with initial conditions $\lambda_1^+(0)=0.9$, $\lambda_1^-(0)=0.1$ and $\lambda_2^\pm(0)=0$.}
\end{center}
\end{figure}

We can now analyze the time behavior of the total and classical trace distance correlations described in Sec.~\ref{sec:TotClassCorr}, for the two models studied above. Fig.~\ref{fig:Corr12} and Fig.~\ref{fig:CorrREF} show the dynamics of total $T_\mathrm{TD}$, classical $C_\mathrm{TD}$ and quantum $D_\mathrm{TD}$ correlations quantified by the trace distance for the local phase-flip channels and for the random external fields, respectively. There are several features of note, both in commonality and contrast with the relative entropy distance measure of discord.

Similarly to the case of relative entropy distance, the trace distance classical correlations switch between being frozen and varying at exactly the same points in time as the trace distance discord. This behavior, known as sudden transition between classical and quantum decoherence \cite{mazzola2010PRL}, can be understood from the analytic expressions of Eqs.~(\ref{TDDiscordfinal}) and (\ref{TDClassicalfinal}), from which we can see that, for trace distance, quantum correlations depend only on $R_{\mathrm{int}}$ for Bell diagonal states, whereas classical correlations depend only on $R_{\max}$. While this is not true in general for the relative entropy distance [see Eqs.~(\ref{entropicquantifiers})], this turns out to be the case for trajectories which experience frozen entropic discord \cite{mazzola2010frozen}. Similarly, for both measures, the total correlations do not appear to experience freezing or sudden change, indicating their dependence on more than one $R_{ii}$ value.

In contrast to the relative entropy distance, however, where $C(t^{\star}) = D(t^{\star})$ at any threshold time $t^{\star}$ at which there is a sudden change, for trace distance $C_\mathrm{TD}(t^{\star}) < D_\mathrm{TD}(t^{\star})$. This too can be understood by reference to the analytic expressions, which show that $C_\mathrm{TD} <  D_\mathrm{TD}$ whenever $R_{\max} = R_{\mathrm{int}}$.

\subsection{Scaling of the freezing regions of quantum correlations}
We now show a general scaling property  of the freezing region for quantifiers of quantum correlations as a function of the initial conditions. This property, which is found for local Markovian nondissipative channels \cite{aaronsonPRA}, can be generalized to any local channel maintaining the Bell diagonal structure of the two-qubit density matrix with $R_{ii}(t)=R_{ii}(0)f^2(t)$, $R_{jj}(t)=R_{jj}(0)f^2(t)$ and $R_{kk}(t)=R_{kk}(0)$ ($i,j,k=1,2,3$, $i,j\neq k$), where $f(t)$
is a characteristic time-dependent function of the channel with the properties $f(0)=1$ and $|f(t)|\leq1$. In fact, the initial conditions for general freezing are $R_{ii}(0)=\pm 1$, $R_{jj}(0)=\mp R_{kk}(0)$ \cite{aaronsonPRA}. Assuming that $|R_{ii}(t)|,|R_{kk}(t)|\geq |R_{jj}(t)|$ for any $t$, the freezing occurs when $|R_{ii}(t)|\geq |R_{kk}(t)|=|R_{kk}(0)$, that is when $f^2(t)\geq |R_{kk}(0)|$. If the function $f(t)$ is analytically invertible, the threshold times $t^\star$ when there is a sudden change can be explicitly determined from the equation $f^2(t^\star)=|R_{kk}(0)|$. Due to the properties of $f(t)$, the general result under these conditions is thus that the smaller is $|R_{kk}(0)|$, the longer is the freezing region of quantum correlations whose amount however correspondingly decreases.

\begin{figure}[tb]
\begin{center}
{\includegraphics[width=0.53\textwidth]{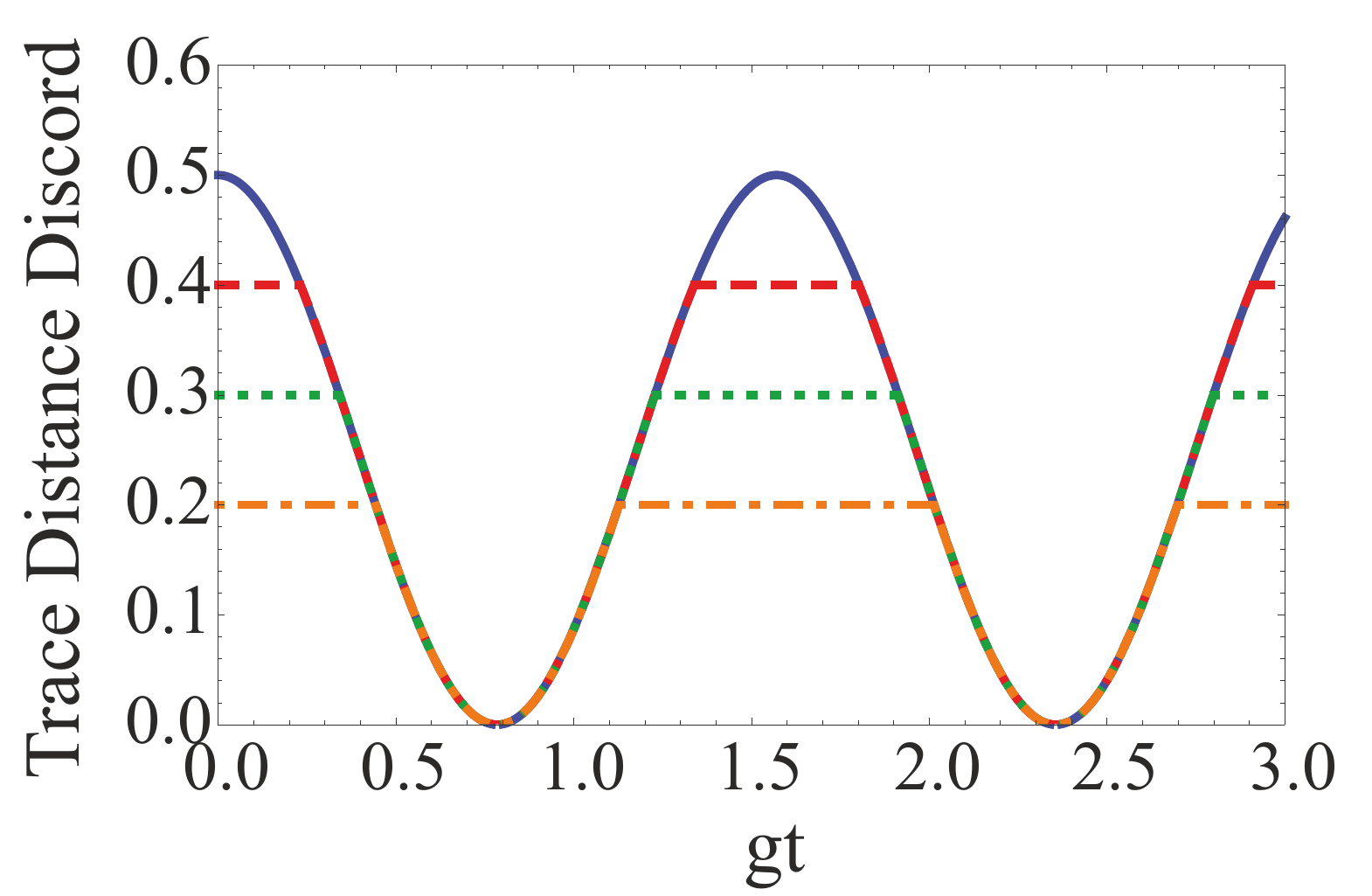}}
\caption{\label{fig:freezingscaling}(Color online) Dynamics of trace distance discord $D_\mathrm{TD}$ for an initial Bell diagonal state with $\lambda_2^\pm(0)=0$ ($R_{33}(0)=-1$) and different values of $\lambda_1^+(0)=1-\lambda_1^-(0)$ given by: $\lambda_1^+(0)=1$, that is $R_{11}(0)=R_{22}(0)=1$ (blue solid line); $\lambda_1^+(0)=0.9$, that is $R_{11}(0)=R_{22}(0)=0.8$ (red dashed line); $\lambda_1^+(0)=0.8$, that is $R_{11}(0)=R_{22}(0)=0.6$ (green dotted line); $\lambda_1^+(0)=0.7$, that is $R_{11}(0)=R_{22}(0)=0.4$ (orange dot-dashed line).}
\end{center}
\end{figure}

For example, in the case of local random external fields considered above we find that the general freezing of quantum correlations occurs when $\cos^2(2gt)\geq |R_{22}(0)|$ and the first sudden change time is at $g t^\star=\frac{1}{2}\arccos\sqrt{|R_{22}(0)|}$. In Fig.~\ref{fig:freezingscaling} we display the scaling of freezing by plotting the trace distance discord as a function of the dimensionless time $gt$ for different values of the initial coefficients, fixing $\lambda_2^\pm(0)=0$ (i.e., $R_{33}=-1$). We notice that by decreasing the value of $\lambda_1^+$ (therefore of $|R_{22}(0)|$), the regions of freezing become longer and the amount of preserved quantum correlations smaller. This phenomenon is universal among all {\it bona fide} measures of quantum correlations as a consequence of the analysis in Ref.~\cite{aaronsonPRA}. For trace distance discord this scaling of the freezing regions can furthermore occur also with initial conditions outside those for general freezing (for instance, for $|R_{33}|\neq1$, see Right panel of Fig.~\ref{fig:DREF}).

\section{\label{sec:Conc}Conclusion}

Bell diagonal states of two qubits are often the simplest yet highly relevant class of states for which one is able to analytically calculate measures of correlations. Investigations of different types of correlations in Bell diagonal states can reveal insights into remarkable dynamical features such as frozen quantum correlations \cite{mazzola2010PRL,aaronsonPRA}, and can lead to a deep understanding of the structure and interplay of different forms of (non)classical correlations.

In this paper, we adopted the trace distance as a metric to define correlations in bipartite quantum states. Extending the analysis of \cite{noq,sarandy} in which a discord measure based on trace distance was defined, we completed a unified approach to bipartite correlations by defining classical and total correlations based on the trace distance metric. For Bell diagonal states, we obtained analytical expressions for classical and total trace distance correlations, in addition to the known one for quantum correlations \cite{noq,sarandy}. Interestingly, trace distance discord is entirely specified by the intermediate Bloch correlation element of Bell diagonal states, while trace distance classical correlations only depend on the maximum Bloch correlation element for the same states. The total correlations have a nontrivial expression which depends on all the Bloch elements, and are obtained for a state $\rho$ by taking the trace distance from a product state which is not, in general, equal to the product of the marginals of $\rho$.

This is an interesting fact in its own right, which did not seem to be noticed before: the product state $\pi$ minimizing the trace distance, i.e.~the probability of error in discriminating $\pi$ from the correlated state $\rho$, is not the product of the marginals of $\rho$ in general. We presented explicit examples including Werner states and rank-$2$ Bell diagonal states, where this fact became manifest. Unlike relative entropy-based approaches to correlations \cite{Modi2010PRL}, for trace distance the total correlations are almost never equal to the sum of classical and quantum ones, but stay strictly smaller than that.

We have examined the behavior of quantum, classical, and total trace distance correlations in two  simple non-Markovian dynamical models: qubits under local phase-flip channels, and under the action of a random external field. The sudden transition between classical and quantum decoherence, first demonstrated for entropic quantifiers of correlations \cite{mazzola2010PRL}, occurs as well for trace distance correlations. However, the trace distance measures exhibit unique  qualitative features, including the presence of frozen discord \cite{aaronsonPRA} under a greater range of starting states compared to other measures of quantum correlations.

The simple expressions obtained in this paper for trace distance correlations of Bell diagonal states make them amenable to precise experimental verification in highly controllable dynamical implementations realized either with photons \cite{can-guo2010NatComm,sciar} or with nuclear magnetic resonance techniques  \cite{isabela}. It might be intriguing to investigate in the future whether the gap between the trace distance total correlations and the sum of trace distance classical correlations plus discord can be of any operational significance in some information processing task. To our knowledge, one operational interpretation for a trace distance based quantifier of correlations was reported for the trace distance discord in the context of remote state preparation fidelity for noisy one-way quantum computations \cite{chavesito}. More generally, we reiterate that the trace distance discord also quantifies operationally the minimum entanglement (negativity) activated between a two-qubit system and an apparatus during a local premeasurement \cite{acti,streltsov}, as very recently observed experimentally \cite{sciar}.

Another interesting direction for future investigation would be to add one more layer to the hierarchy of trace distance correlations by computing the minimum distance from the set of separable states, which would define a measure of entanglement \cite{horodecki2009RMP} based on trace distance. Finally, we can expect that some of the results presented here can be extended to more general classes of two-qubit states such as the $X$-shaped density matrices, adopting the methods of \cite{giovannetti}.

\medskip

\noindent {\it Note added.}--- After completion of this work we became aware of similar results obtained independently by F. M. Paula {\it et al.} \cite{sarandynew}. However, there the authors define trace distance quantifiers of classical and total correlations for a bipartite state $\rho_{AB}$ without including a minimization over the closest product state, but considering distances from a fixed reference product state given by the product of the marginals of $\rho_{AB}$. As a result, their quantities exceed ours and their definitions might overestimate the content of correlations in quantum states, especially in the case of the total trace distance correlations. Interestingly, the classical trace distance correlations according to both their definition and our optimized one, turn out to be monotonic functions of each other for Bell diagonal states, being both dependent only on the maximum correlation element $R_{\max}$.

\section*{Acknowledgments}
We are grateful to I. A. Silva, D. Soares-Pinto, and M. S. Sarandy  for fruitful discussions.
We acknowledge financial support from the University of Nottingham through an Early Career Research and Knowledge Transfer Award and an EPSRC Research Development Fund Grant (PP-0313/36), and from the Brazilian funding agency CAPES [Pesquisador Visitante Especial-Grant No. 108/2012].

\section*{References}

\bibliographystyle{iopart-num}

\providecommand{\newblock}{}

\end{document}